\documentclass[preprint,12pt]{elsarticle}
\usepackage{amsmath}
\usepackage{amssymb}
\usepackage{graphicx}
\usepackage{hyperref}
\usepackage{times}

\newcommand{\dopreprint}{0}

\ifcase\dopreprint
\newcommand{\fwidth}{.45\textwidth}  
\else
\newcommand{\fwidth}{.75\textwidth}  
\fi

\journal{Physics Letters A}

\begin{document} 

\begin{frontmatter}

\title{Robust chaos with prescribed natural invariant measure and Lyapunov exponent}
\author{Juan~M.~Aguirregabiria\fnref{em}}
\address{Theoretical Physics, 
The University of the Basque Country, \\
P.~O.~Box 644,
48080 Bilbao, Spain}
\fntext[em]{Corresponding author: juanmari.aguirregabiria@ehu.es
\\Tel. +34 946105915, Fax +34 946013500}

\date{\today}

\begin{abstract} 
We extend in several ways a recently proposed method to construct one-dimensional
chaotic maps with exactly known natural invariant measure 
\cite{Sogo1,Sogo2}. First, we assume that the given invariant measure depends on a continuous parameter and
show how to construct maps with robust chaos ---i.e., chaos that is not destroyed by arbitrarily small changes
of the parameter--- and prescribed invariant measure and constant Lyapunov exponent. 
Then, by relaxing one condition in the approach of Refs. \cite{Sogo1,Sogo2}, we 
describe a  method to construct robust chaos with prescribed constant invariant measure and 
varying Lyapunov exponent. 
Another extension of a condition in Refs.\ \cite{Sogo1,Sogo2} provides a new method to get robust chaos with
known varying Lyapunov exponent.
In this third approach the
invariant measure can be computed exactly in many particular cases. Finally we discuss how to use diffeomorphisms to
construct maps with robust chaos, any number of parameters and prescribed invariant measure and Lyapunov exponent.
\end{abstract} 

\begin{keyword}
nonlinear dynamical system\sep  deterministic chaos\sep  robust chaos\sep natural invariant measure
\PACS 05.45.Ac\sep 05.45.-a
\end{keyword}

\end{frontmatter}

\section{\label{sec:intro}Introduction}

The inverse problem for chaotic one-dimensional maps has been recently proposed and
solved in some cases by Sogo \cite{Sogo1,Sogo2}. Starting from a given invariant measure,
the method allows to construct maps with that measure and Lyapunov exponents
in the form $\lambda=\ln m$, with $m=2,3,\ldots$ This method can be used to construct
exact examples of chaotic one-dimensional maps 
depending on the discrete parameter $m$ and having exactly known invariant measure and Lyapunov exponent.
The first goal of this work is to extend this method to families
of maps that depend on a continuous parameter, display robust chaos
---i.e., they have a 
chaotic attractor which is not destroyed by arbitrarily small changes of the parameter---
and have known invariant measure and Lyapunov exponent.

Piecewise smooth maps may show robust chaos and
have been used to describe robust chaos in circuits \cite{Banerjee}.
On the other hand, many families of smooth maps have fragile chaos. 
For instance,
the logistic map,
$x_{n+1}=4r x_n\left(1-x_n\right)$, is chaotic for $r=1$, 
but the attractor is periodic for a set of values of the parameter $r$ that is
dense in the interval $0\le r\le 1$ \cite{Ott}. If such a dynamical system
describes a real device, it may be impossible to know
in advance whether the behavior of the system
will in fact be chaotic or periodic for some parameter value, which is always known
with finite precision. However, robust chaotic behavior is required in
many applications, including 
encrypting messages \cite{Hayes,Vaidya}, 
 random number generators \cite{Galajda}
and  engineering applications \cite{Elhadj}.
Robust chaos also arises in neural networks \cite{Priel,Potapov} as well as 
in the study of brain  and
population dynamics \cite{Dafilis,Ros}. 
Thus, simple and easy ways of constructing robust chaotic attractors with
exactly known properties, 
as those explored in this work,
can be of interest in different fields of science and engineering. 

Andrecut and Ali first found a smooth map
\cite{Andrecut1} and later a method of generating smooth maps \cite{Andrecut2} whose  evolution is
chaotic for whole intervals of the parameter. Another case is discussed in Ref. \cite{Shastry} and we have
recently  explored several new ways to construct smooth
one-dimensional maps with robust chaos \cite{JMA}.

The purpose of this work is to present other ways of constructing maps with robust chaos.
The advantage of these new methods is that the Lyapunov exponent is
known exactly and the invariant measure is known always with the first, second and fourth methods and in many
particular cases in the third approach. The first three methods will be extensions of Sogo's
inverse problem \cite{Sogo1,Sogo2}. In Sect.\ \ref{sec:robust} we will
show how to construct families of chaotic maps with a constant Lyapunov exponent
and a given invariant measure
depending continuously on a parameter. 
In Sect.\ \ref{sec:constant} we present a method to construct families of chaotic
maps sharing the same prescribed invariant measure and having a known Lyapunov
exponent depending on the parameter.
In
Sect.\ \ref{sec:gen} we will explore a method to construct robust chaos with a
known Lyapunov exponent varying continuously with the parameter and a natural measure
that can be computed exactly in many particular cases. Finally, in Sect.\ \ref{sec:diffeo} and \ref{sec:inverse} 
we will discuss several ways to construct robust chaos with prescribed Lyapunov exponent and natural 
invariant measure by using diffeomorphisms.

We will consider one-dimensional maps on a finite interval $[a,b]$, which for commodity will be reduced to $[0,1]$
 by means of a linear transformation.

\section{\label{sec:robust}Robust chaos with prescribed invariant measure}

We will use an easy extension of Sogo's method \cite{Sogo1,Sogo2} to construct families
of dynamical systems, $x_{n+1}=f_r\left(x_n\right)$, that are chaotic for a full
range of the parameter $r$ and have exactly known invariant measure and Lyapunov exponent.
Let us assume that the maps $f_r$ are $m$-to-1 so that each value $x\in[0,1]$ has $m$
preimages $y_k$, such that $f_r\left(y_k\right)=x$, for $k=1,2,\ldots,m$. 
the natural invariant measure $d\mu_r=\rho_r(x)\,dx$ satisfies the Frobenius-Perron
equation \cite{Ott}:
\begin{equation}\label{eq:FP}
\rho_r(x)=\int_0^1 \rho_r(y)\delta\left(x-f_r(y)\right)\,dy=\sum_{k=1}^m\frac{\rho_r\left(y_k\right)}{|f'_r\left(y_k\right)|}.
\end{equation}
To find solutions of this equation we will simplify it by 
further assuming that all the terms in the sum make the same contribution, so that
the substitution $y_1\to x$ gives the condition
\begin{equation}\label{eq:eqdifm}
\rho_r\left(f_r(x)\right)\,|f'_r(x)|=m\rho_r(x).
\end{equation}

Eq.\  (\ref{eq:eqdifm}) provides a useful practical way to construct a family of maps $f_r$ 
with a prescribed invariant density $\rho_r$: just choose, for a given integer value of $m\ge2$, a 
family of solutions
of the differential equation (\ref{eq:eqdifm}) that map the phase space $[0,1]$ onto itself and are $m$-to-1. 
This will be a family of maps with robust chaos. By extending the argument in Ref.\ \cite{Sogo2}
one can show that all the maps in the family will have the same Lyapunov exponent:
\begin{align}\label{eq:lambdam}
\lambda_r&=\int_0^1\rho_r(x)\ln\left|f'_r(x)\right|\,dx \nonumber\\&=
\ln m+\int_0^1 \rho_r(x)\ln \rho_r(x)\,dx
\nonumber\\&\qquad-\int_0^1 \rho_r(x)\ln \rho_r\left(f_r(x)\right)\,dx=\ln m.
\end{align}
The two last integrals cancels each other because of (\ref{eq:FP}):
\begin{multline}
\int_0^1 \rho_r(x)\ln \rho_r(x)\,dx\\
=\int_0^1dy\,\rho_r(y)\int_0^1dx\,\delta\left(x-f_r(y)\right)\ln \rho_r(x)\\
=\int_0^1 \rho_r(y)\ln \rho_r\left(f_r(y)\right)\,dy.
\end{multline}

\subsection{Some solutions with $m$=2}
To give some examples, let us further assume $m=2$ and that the invariant densities $\rho_r\in C^0(0,1)$
are strictly positive, $\rho_r(x)>0$ for $0<x<1$, so that
\begin{equation}\label{eq:mu}
\mu_r(x)\equiv\int_0^x\rho_r(y)\,dy
\end{equation}
monotonously increases from $\mu_r(0)=0$ to $\mu_r(1)=1$ and has a unique
inverse $\mu_r^{-1}(x)$ in $[0,1]$.
If we choose the boundary conditions $f_r(0)=f_r(1)=0$, the solution of (\ref{eq:eqdifm})
is
\begin{equation}\label{eq:solmu1}
f_r(x)=\mu_r^{-1}\left(1-\left|1-2\mu_r(x)\right|\right).
\end{equation}
Each map will  increase from $f_r(0)=0$ to the maximum
\begin{equation}
f_r\left(\alpha_r\right)=1,\qquad \alpha_r\equiv\mu_r^{-1}\left(\frac12\right)
\end{equation}
and then decrease to $f_r(1)=0$. The Lyapunov exponent will be $\lambda_r=\ln2$.

\subsubsection{A family of piecewise smooth chaotic maps}
If one chooses the invariant densities
\begin{equation}\label{eq:rhopot}
\rho_r(x)=rx^{r-1},\quad(r>0),
\end{equation}
the solution (\ref{eq:solmu1}) will be
\begin{equation}\label{eq:fam1a}
f_r(x)=\left(1-\left|1-2x^r\right|\right)^{1/r},\quad (r>0).
\end{equation}
Three members of this family of chaotic maps are displayed in 
Fig.\ \ref{fig1}. The full tent map $T(x)=1-|1-2x|$ is recovered with $r=1$.
\ifcase \dopreprint
\begin{figure}
\begin{center}
\includegraphics[width=\fwidth]{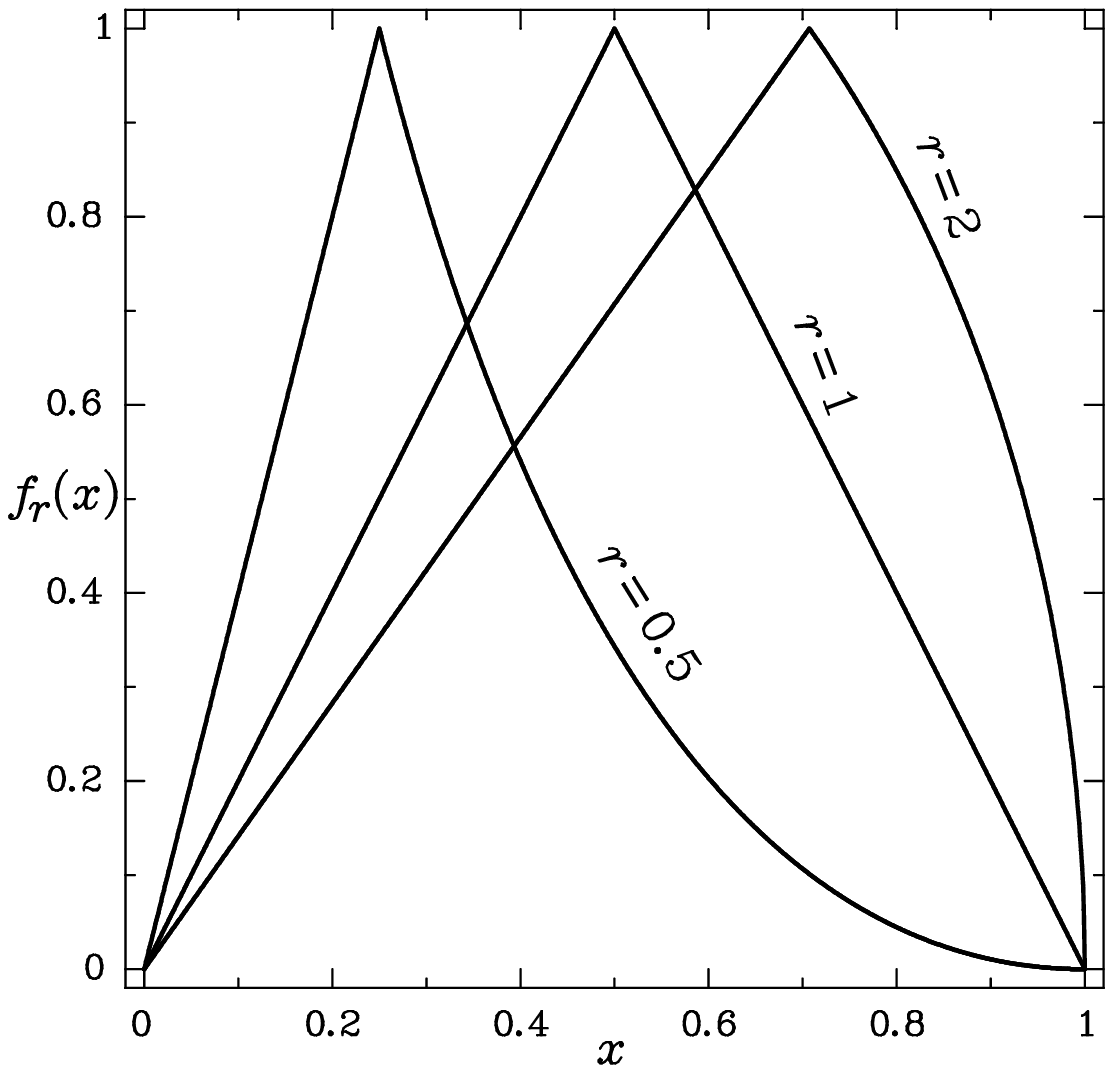}
\end{center}
\caption{Three members of the family of maps (\ref{eq:fam1a}).\label{fig1}} 
\end{figure}
\fi

\subsubsection{A family of smooth chaotic maps}
The solutions (\ref{eq:fam1a}) are not differentiable at $x=\alpha_r$. In fact,
since
\begin{equation}
\lim_{x\to\alpha_r^\pm}f'_r(x)=\mp\frac{2\rho_r\left(\alpha_r\right)}{\rho_r(1)},
\end{equation}
the condition for $f'_r$ to be continuous at the maximum is $\lim_{x\to1}\rho_r(x)=\infty$.

We can use this condition to find families of chaotic maps that are smooth along the
full interval $[0,1]$. To provide an example, 
let us consider the family of densities given for every $r>0$ by  
\begin{equation}
  \rho_r(x)=\frac{2^{r-1}r\arcsin^{r-1}\sqrt x}{\pi^r\sqrt{x(1-x)}}.
\end{equation}
Then  the solution (\ref{eq:solmu1}) is
\begin{equation}\label{eq:fam2}
 f_r(x)=
\begin{cases}\sin^2\left(2^{1/r}\arcsin\sqrt x\right),&\displaystyle0\le x\le \alpha_r;\\
             \sin^2\left(2^{1/r}\left(\frac{\pi^r}{2^r}-\arcsin^r\sqrt x\right)^{1/r}\right),
             &\alpha_r\le x\le1,
\end{cases}
\end{equation}
with $\alpha_r\equiv \sin^2\left(2^{-(1+1/r)}\pi\right)$. 
\ifcase \dopreprint
\begin{figure}
\begin{center}
\includegraphics[width=\fwidth]{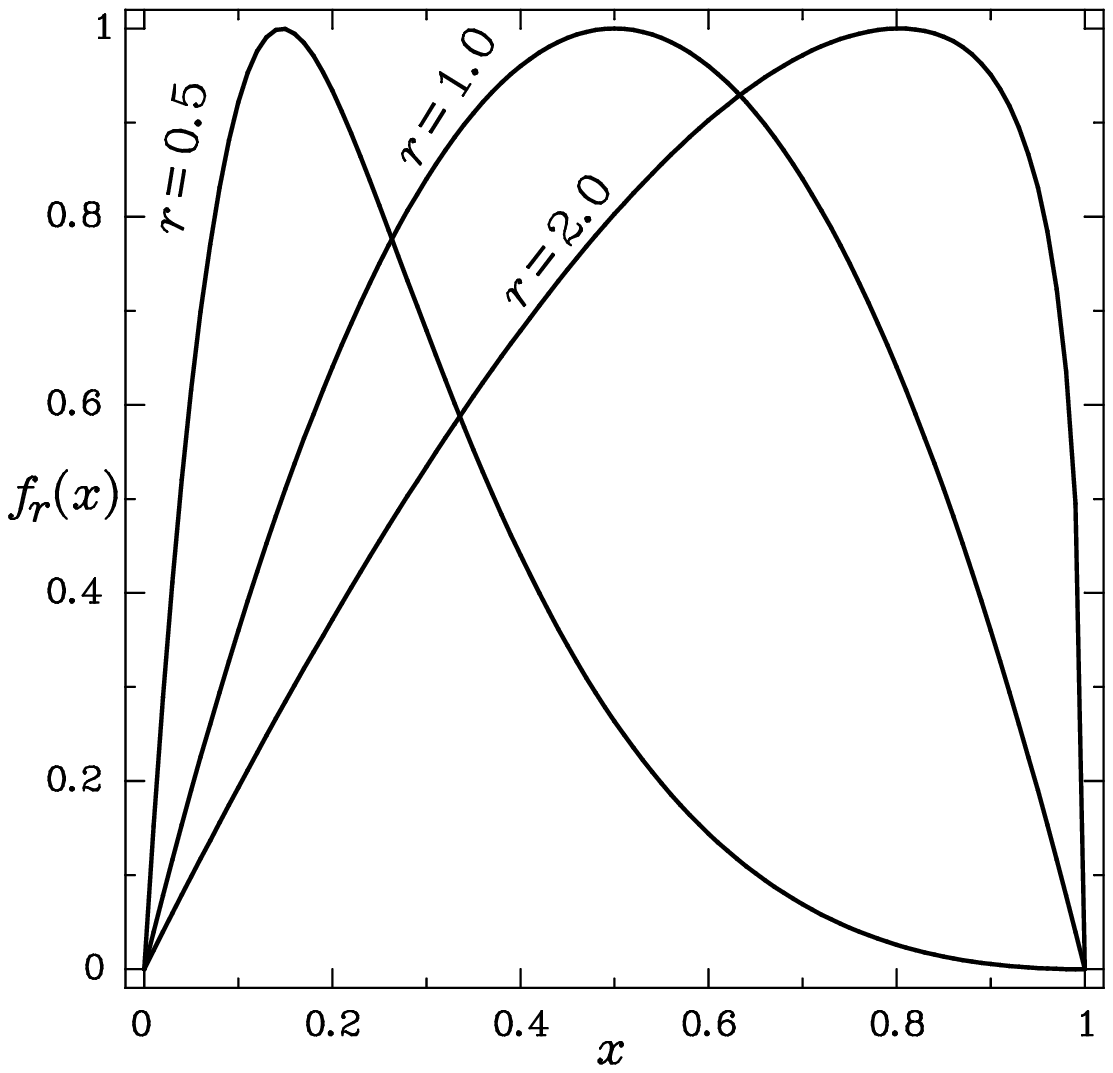}
\end{center}
\caption{Three members of the family of maps (\ref{eq:fam2}).\label{fig2}} 
\end{figure}
\fi

Some members of the family (\ref{eq:fam2}) are displayed in Fig.\ \ref{fig2}, where one recognizes
the full logistic map $f_1(x)=4x(1-x)$. We have $f_r\in C^1[0,1]$ for $0<r\le1$, but $f'_r(x)\to-\infty$ as $x\to 1$
for $r>1$.  

The boundary conditions for $f_r(x)$ can be chosen in many other ways. For instance,
with $f_r(0)=f_r(1)=1$ the maps will have a single minimum, instead of a maximum, and the substitute
for (\ref{eq:solmu1}) is
\begin{equation}\label{eq:solmu1b}
f_r(x)=\mu_r^{-1}\left(\left|1-2\mu_r(x)\right|\right).
\end{equation}
The single minimum  is $f_r\left(\alpha_r\right)=0$ 
with $\alpha_r\equiv\mu_r^{-1}\left(1/2\right)$.

With the densities (\ref{eq:rhopot}) one gets
\begin{equation}\label{eq:fam1}
f_r(x)=\left|1-2x^r\right|^{1/r},\quad (r>0),
\end{equation}
which has a smooth minimum only for $0<r<1$, since now the condition is $\lim_{x\to0}\rho_r(x)=\infty$.

It is also easy to extend the results of this section
for values of $m\ne2$. But let us present another way to construct robust chaos.

\section{\label{sec:constant}Robust chaos with constant invariant measure}

Following Refs.\ \cite{Sogo1,Sogo2}, to get Eq.\  (\ref{eq:eqdifm}) we assumed that all terms in sum (\ref{eq:FP}) had the same
value. It is obvious that this simplifying condition can be relaxed in many ways. Let us explore 
a simple one. One could construct a family of 2-to-1 maps, with the same prescribed invariant density $\rho(x)$
but with a generalized relation between the terms in sum (\ref{eq:FP}),
\begin{equation}
\frac{\rho\left(y_1\right)}{|f'_r\left(y_1\right)|}=r\frac{\rho\left(y_2\right)}{|f'_r\left(y_2\right)|},\quad(r>0),
\end{equation}
for instance. With this assumption, one would replace the constant multiplicity $m$ of Eq.\  (\ref{eq:eqdifm})
by 
\begin{equation}
m_r(x)=
\begin{cases}
 \frac{1+r}r,&0\le x<\alpha_r;\\
 1+r,&\alpha_r<x\le1;
\end{cases}
\end{equation}
where $x =\alpha_r$ is the value where the two preimages coincide: $y_1\left(\alpha_r\right)=y_2\left(\alpha_r\right)$.
The family $f_r$ would satisfy the following equation:
\begin{equation}\label{eq:eqdifmr}
\rho\left(f_r(x)\right)\,|f'_r(x)|=m_r(x)\rho(x), \quad(r>0).
\end{equation}

For example, for any $\rho\in C^0(0,1)$ with $\rho(x)>0$ and 
$\mu(x)\equiv\int_0^x\rho(y)\,dy)$, 
the solution of (\ref{eq:eqdifmr}) with $f_r(0)=f_r(1)=0$ is
\begin{equation}\label{eq:solmu2}
f_r(x)=
\begin{cases}
\mu^{-1}\left(\frac{1+r}r\,\mu(x)\right),&\displaystyle0\le x\le \alpha_r;\\
\mu^{-1}\left((1+r)\left(1-\mu(x)\right)\right),&\displaystyle\alpha_r\le x\le 1;
\end{cases}
\end{equation}
with
\begin{equation}\label{eq:alpharr}
\alpha_r\equiv\mu_r^{-1}\left(\frac r{1+r}\right).
\end{equation}
By using (\ref{eq:eqdifmr}) and (\ref{eq:alpharr}) the varying Lyapunov exponent is readily computed:
\begin{equation}\label{eq:lyap1b}
\lambda_r=\int_0^1\rho(x)\ln\left|f'_r(x)\right|\,dx= \ln(1+r)-\frac r{1+r}\ln r.
\end{equation}
This value starts from $\lim_{r\to 0}{\lambda_r=0}$, increases until its maximum $\lambda_1=\ln 2$ 
and then decreases towards  $\lim_{r\to \infty}{\lambda_r=0}$. The
condition for the maximum at $x=\alpha_r$ to be smooth is again $\lim_{x\to1}\rho(x)=\infty$.

\subsection{Another family of piecewise-smooth chaotic maps}
For example, if one selects for all $r>0$ the same constant invariant density
$\rho_r(x)=\rho(x)=1$ the solution of Eq.\ (\ref{eq:solmu2}) is
the family
\begin{equation}\label{eq:fam1b}
f_r(x)=  (1+r)
\begin{cases}
x/r,&0\le x\le \alpha_r;\\
1-x,&\alpha_r\le x\le1;
\end{cases}\quad (r>0),
\end{equation}
with $\alpha_r=r/(1+r)$
and the Lyapunov exponent of Eq.\ (\ref{eq:lyap1b}).
 $f_1$ is 
the full tent map $T(x)=1-|1-2x|$. 

\subsection{A second family of smooth chaotic maps}
If the starting point is the invariant density of the full logistic map, 
$\rho_r(x)=\rho(x)=\left[\pi^2x(1-x)\right]^{-1/2}$,
we get, for every $r>0$,
\begin{equation}\label{eq:fam1c}
f_r(x)= 
\begin{cases}
\sin^2\left(\frac{1+r}r\arcsin\sqrt x\right),&0\le x\le \alpha_r;\\
\sin^2\left((1+r)\arccos\sqrt x\right),&\alpha_r\le x\le1;
\end{cases}
\end{equation}
with 
\begin{equation}
\alpha_r=\cos^2\frac\pi{2(1+r)}
\end{equation}
and the Lyapunov exponent (\ref{eq:lyap1b}).
The full logistic map is recovered with $r=1$: $f_1(x)=4x(1-x)$.

It is straightforward to extend the method in this section for $m=3,4,\ldots$

\section{\label{sec:gen}Robust chaos with varying Lyapunov exponent}
In Eq.\  (\ref{eq:eqdifm}) it was assumed that $m$ is an integer larger than one. But let us substitute
for it a  real number $r>1$:
\begin{equation}\label{eq:eqdifr}
\rho\left(f_r(x)\right)\,|f'_r(x)|=r\rho(x), \quad(r>1).
\end{equation}
Notice that  we are using again the same invariant density $\rho(x)$ for all values of the continuous 
parameter $r$, which now replaces the constant multiplicity $m$.
If we get a solution of Eq. (\ref{eq:eqdifr}) for a given $\rho(x)$,
the latter will not satisfy the Frobenius-Perron equation for non-integer values of $r$, but this does
not prevent the solution $f_r$ from being chaotic for all $r>1$, as we will show in the following.
In fact, by using an argument very similar to the one leading to (\ref{eq:lambdam}), one can show that
the Lyapunov exponent of the family satisfying (\ref{eq:eqdifr})
will be $\lambda_r=\ln r$, for all $r>1$.

The solution of Eq.\ (\ref{eq:fam3}) for $f_r(0)=0$, $f'_r(0)>0$ and $1<r\le2$ can be written
as
\begin{equation}\label{eq:solfam3}
f_r(x)=
\begin{cases}
 \mu^{-1}\left(r\mu(x)\right),&0\le x\le \alpha_r\equiv\mu^{-1}(1/r);\\
 \mu^{-1}\left(2-r\mu(x)\right),&\alpha_r\le x\le1,
\end{cases}
\end{equation}
with $\mu(x)\equiv\int_0^x\rho(y)\,dy$. It is also easy to write down the solution 
for other initial conditions or
other ranges of the parameter $r$.

\ifcase \dopreprint
\begin{figure}
\begin{center}
\includegraphics[width=\fwidth]{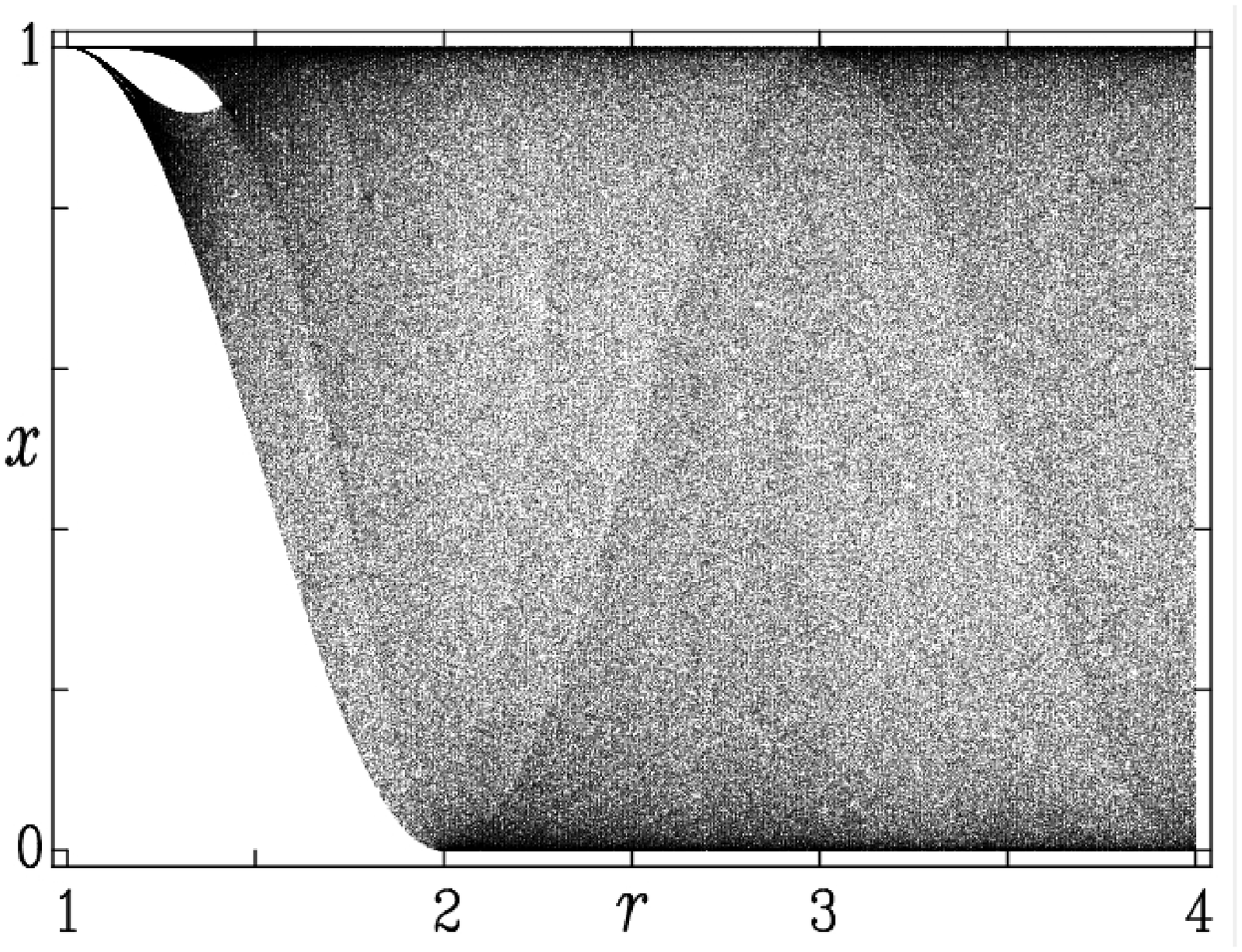}
\end{center}
\caption{Bifurcation diagram of the map (\ref{eq:fam3}).\label{fig3}} 
\end{figure}
\fi
\subsection{A third family of smooth chaotic maps}
If we choose again the invariant density of the full logistic map, $\rho(x)= [\pi^2x(1-x)]^{-1/2}$, 
a solution of Eq.\ (\ref{eq:eqdifr}) for all $r>1$ is
\begin{equation}\label{eq:fam3}
f_r(x)=\sin^2\left(r\arcsin\sqrt x\right).
\end{equation}
For $r=2,3,\ldots$ the maps $f_r(x)$ 
reduce to polynomials and we recover the `Chebyshev hierarchy' of Refs. \cite{Sogo1,Sogo2}:
\begin{equation}
f_n(x)=\frac12-\frac{(-1)^n}2\,T_n(2x-1),\quad(n=2,3,\ldots),
\end{equation}
where $T_n$ are the Chebyshev polynomials of the first kind. All these maps are chaotic, with Lyapunov exponent
$\lambda_n=\ln n$ and the same invariant invariant density as the full logistic map
$f_2(x)=4x(1-x)$. The question is what happens for non-integer values of $r$?

Their Lyapunov exponent is $\lambda_r=\ln r$ and we can see in the bifurcation diagram of Fig.\ \ref{fig3} 
that 
the attractor only fills the phase space $[0,1]$ for $r\ge2$.

But, what about the invariant measure? The point is that now
all points $x$ do not have the same number $m$ of preimages $y_k$, such that $f_r\left(y_k\right)=x$, as assumed
in Refs. \cite{Sogo1,Sogo2} and Sect.\ \ref{sec:robust} and \ref{sec:constant}. In consequence, the starting
invariant density will not satisfy the Frobenius-Perron equation
(\ref{eq:FP}) for non-integer values of $r$. 
Obviously, 
the actual  invariant measure will satisfy that equation with a
value of $m$ depending on the point $x$. It it clear from Fig.\ \ref{fig4}, which display in solid line the
graphs of two members of family (\ref{eq:fam3}),
that the value $m(x)$ will change in this example at point
$x=\alpha\equiv f_r(1)$ and the same will happen  at all its images, so that we can expect the invariant measure to
be discontinuous at every point in the orbit of $f_r(1)$:
\begin{equation}\label{eq:orbit1}
\mathcal O_r\equiv\left\{f_r^k(1):\ k=1,2,\ldots\right\}.
\end{equation}
\ifcase \dopreprint
\begin{figure}
\begin{center}
\includegraphics[width=\fwidth]{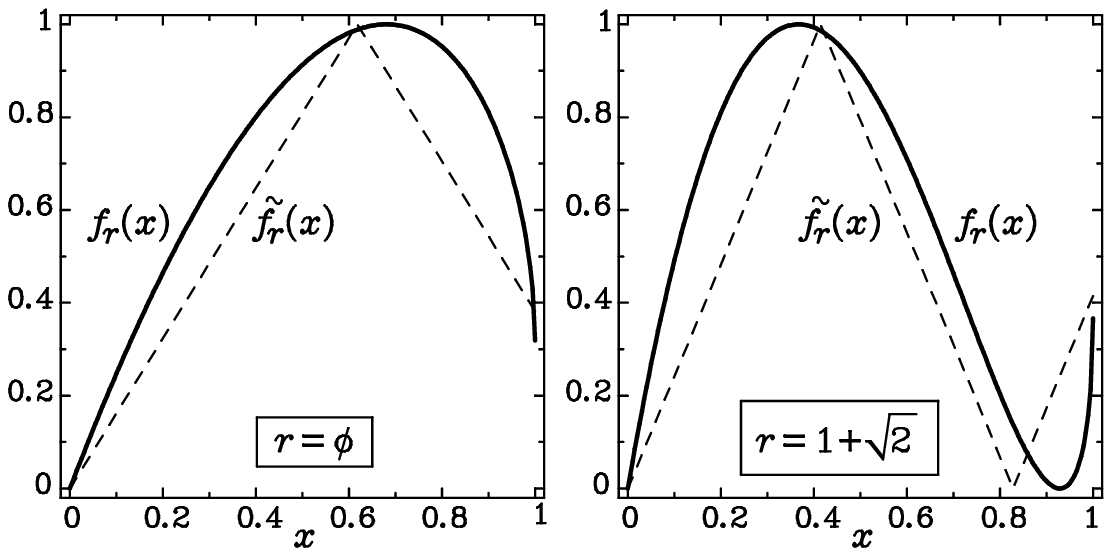}
\end{center}
\caption{The maps (\ref{eq:fam3})  and (\ref{eq:saw}) for $r=\phi,\ 1+\sqrt2$.\label{fig4}} 
\end{figure}
\fi

The task of finding the invariant measure can thus be very difficult unless the orbit $\mathcal O_r$
is simple enough (a cycle), as happens in the `Chebyshev hierarchy' (where $r=2,3,\ldots$ and $\mathcal O_{2n}=\{0,0,\ldots\}$
and $\mathcal O_{2n-1}=\{1,1,\ldots\}$, so that the invariant density is continuous for $0<x<1$), 
but also for many non-integer values of $r$
as we shall see now.

For instance, in the left graph of Fig.\ \ref{fig4} the parameter equals the golden ratio, $r=\phi\equiv(1+\sqrt5)/2$, 
and then the orbit of $f_\phi(1)$ is a 3-cycle:
\begin{equation}\label{eq:orbit}
\mathcal O_\phi=\left\{\alpha \equiv\sin^2\frac{\phi\pi}2,\ 1-\alpha,\ 1,\ \alpha,\ 1-\alpha,\ 1,\ldots\right\}.
\end{equation}
Since one can multiply a solution of Eqs. (\ref{eq:FP}) and (\ref{eq:eqdifr}) with any constant,
one may suspect that the actual invariant measure of the maps (\ref{eq:fam3}) is that of the full logistic map
 multiplied by a function which is constant except at the points lying on the orbit of $f_\phi(1)$. 
In fact,
we can find the exact expression of the invariant density for $r=\phi$ 
by using the normalization condition $\int_0^1\rho_\phi(x)\,dx=1$
and trying in the Frobenius-Perron equation (\ref{eq:FP})  a density in the form
\begin{equation}
\rho_\phi(x)=\frac1{\pi\sqrt{x(1-x)}}
\begin{cases}
a,&0< x<\alpha ;\\
b,&\alpha < x<1-\alpha ;\\
c,&1-\alpha < x<1;
\end{cases}
\end{equation}
with constant $a$, $b$ and $c$ and the following multiplicity:
\begin{equation}
m(x)=
\begin{cases}
1,&0< x<\alpha ;\\
2,&\alpha < x<1.
\end{cases}
\end{equation}
One finally finds
\begin{equation}
a=0,\quad b= \frac c\phi=\frac{1+3\phi}5.
\end{equation}
The result can be easily checked by means of numerical simulations \cite{DS}.

Even simpler is the orbit of $f_r(1)$ for $r=1+\sqrt2$:
\begin{equation}
\mathcal O_{1+\sqrt2}=\left\{\alpha =\cos^2\frac{\pi}{\sqrt2},\ 1,\ \alpha ,\ 1,\ldots\right\}.
\end{equation}
From this 2-cycle and the Frobenius-Perron equation (\ref{eq:FP}) it is easy to find the invariant  density:
\begin{equation}\label{eq:sqrt2}
\rho_{1+\sqrt2}=\frac{2+\sqrt2}{4\pi\sqrt{x(1-x)}}
\begin{cases}
\sqrt2,&0< x<\alpha ;\\
1,&\alpha < x<1.
\end{cases}
\end{equation}

Other cases with piecewise continuous invariant measure can be found in a similar way.

It should be noticed that maps (\ref{eq:fam3}) are a nice example of solvable robust chaos,
since the solution can be explicitly written in terms
of the initial condition $x_0$ as $x_n=\sin^2\left(r^n\arcsin\sqrt{x_0}\right)$.

\section{\label{sec:diffeo}Robust chaos through diffeomorphisms}

Another way to construct examples of robust chaos
takes advantage of the invariance of Lyapunov exponents through diffeomorphisms
on the interval $[0,1]$. We are going to consider three different approaches.

\subsection{A single diffeomorphism}
If we know a family of chaotic maps $f_r(x)$ with invariant densities
$\rho_r(x)$ and choose a diffeomorphism $\varphi(x)$,
the topologically conjugated family $\tilde f_r=\varphi\circ f_r\circ\varphi^{-1}$ will
have the same Lyapunov exponent and the following invariant density \cite{Ott}:
\begin{equation}\label{eq:rhoconj}
\tilde\rho_r(x)=\frac{\rho_r\left(\varphi^{-1}(x)\right)}{\left|\varphi'\left(\varphi^{-1}(x)\right)\right|}.
\end{equation}

For example if $f_r$ is the family (\ref{eq:fam1b}) and the diffeomorphism
\begin{equation}
\varphi(x)=\sin^2\frac{\pi x}2,
\end{equation}
the topologically conjugated family is (\ref{eq:fam1c}).

On the other hand, if $f_r$ is the family (\ref{eq:fam3}) and the diffeomorphism
\begin{equation}
\varphi(x)=\frac2\pi\arcsin\sqrt x,
\end{equation}
the topologically conjugated family is
\begin{equation}\label{eq:saw}
\tilde f_r(x)=1-\left|1-rx\bmod2\right|.
\end{equation}
Maps (\ref{eq:saw}) are piecewise linear and two particular cases are depicted in dashed line in Fig.\ \ref{fig4}.
Moreover, one recovers the full tent map for $r=2$.
It is obvious that since the slope of the graph of $\tilde f_r(x)$ is nearly everywhere $\pm r$ the 
maps are chaotic for all $r>1$ and that the Lyapunov exponent is $\lambda_r=\ln r$. This in turn proves again that
the topologically conjugated family (\ref{eq:fam2}) is chaotic and has the same Lyapunov coefficient.
The invariant density is $\tilde\rho_n(x)=1$
for $r=n=2,3,\ldots$ and can be computed for other values of $r$ by using (\ref{eq:rhoconj}) or the method discussed
in the last section. For example, the measure corresponding to (\ref{eq:sqrt2}) is
piecewise constant:
\begin{equation}\label{eq:sqrt21}
\rho_{1+\sqrt2}(x)=\frac{2+\sqrt2}{4}
\begin{cases}
\sqrt2,&0\le x<\sqrt2-1 ;\\
1,&\sqrt2-1 < x\le1.
\end{cases}
\end{equation}

\subsection{A family of diffeomorphisms}
A second possibility is to choose a single chaotic map $f(x)$ with known invariant density $\rho(x)$ and
Lyapunov exponent $\lambda$
and a family of diffeomorphisms depending on a continuous parameter: $\varphi_r(x)$. Then
the maps $\tilde f_r=\varphi_r\circ f\circ\varphi_r^{-1}$ will
have the same Lyapunov exponent and the following natural density:
\begin{equation}\label{eq:rhoconjr}
\tilde\rho_r(x)=\frac{\rho\left(\varphi_r^{-1}(x)\right)}{\left|\varphi_r'\left(\varphi_r^{-1}(x)\right)\right|}.
\end{equation}

For instance, if the starting map is the full tent map, $f(x)=T(x)=1-|1-2x|$, 
with $\rho(x)=1$ and $\lambda=\ln 2$ and the diffeomorphism $\varphi_r(x)=x^{1/r}$, with $r>0$, the family
$\tilde f_r$ is precisely (\ref{eq:fam1a}) with the same Lyapunov exponent and the invariant density 
 (\ref{eq:rhopot}).
(Notice that in this example $f=\tilde f_1$.)

On the other hand, given the full tent map 
and the diffeomorphisms $\varphi_r(x)=\sin^{1/r}\frac{\pi x}2$, with $r>0$,
the topologically conjugated smooth maps
\begin{equation}\label{eq:smooth}
\tilde f_r(x)=
\sin^{1/r}\left(2\arcsin x^r\right)=x\left(4-4x^{2r}\right)^{1/2r}
\end{equation}
have the same Lyapunov exponent and the natural densities
\begin{equation}
\tilde\rho_r(x)=\frac{2rx^{r-1}}{\pi\sqrt{1-x^{2r}}}.
\end{equation}
This map, which is also obtained from the full logistic map
$f(x)=4x(1-x)$ by means of the diffeomorphism $\varphi_r(x)=x^{1/2r}$, is a simple example of exactly solvable robust chaos, in which everything is known,
including the general solution $x_n=\sin^{1/r}\left(2^n\arcsin x_0^r\right)$.
We recover the full logistic map with $r=1/2$.

\subsection{Prescribed invariant density}
Finally, one can choose a chaotic map $f(x)$ with known invariant density $\rho(x)$ and
Lyapunov exponent $\lambda$ and a prescribed family of natural densities $\tilde \rho_r(x)$ depending on 
the parameter $r$. Then each
family of diffeomorphisms $\varphi_r$ satisfying the differential equation
\begin{equation}\label{eq:eqdiftil}
 \tilde\rho_r\left(\varphi_r(x)\right)\left|\varphi'_r(x)\right|=\rho(x)
\end{equation}
will provide a family of maps $\tilde f_r=\varphi_r\circ f\circ\varphi_r^{-1}$ with
 the same Lyapunov exponent and the desired natural measure.

If we want $\varphi_r(x)$ to be increasing, the solution of (\ref{eq:eqdiftil}) is
\begin{equation}\label{eq:solphimu}
\varphi_r(x)=\tilde\mu_r^{-1}\left(\mu(x)\right),
\end{equation}
with 
\begin{equation}
\tilde\mu_r(x)\equiv\int_0^x\tilde\rho_r(y)\,dy,\qquad\mu(x)\equiv\int_0^x\rho(y)\,dy.
\end{equation}

For example if $f(x)$ is the full tent map and $\tilde \rho_r(x)=rx^{r-1}$, 
 the solution  (\ref{eq:solphimu}) us $\varphi_r(x)=x^{1/r}$ 
for $r>0$, 
and we recover once more the family (\ref{eq:fam1a}), which in turn shows again
that the latter is chaotic for all positive real values of $r$ and that its invariant density is (\ref{eq:rhopot}).

If we choose $\varphi_r(x)$ to be decreasing, the solution of (\ref{eq:eqdiftil}) is $\varphi_r(x)=\tilde\mu_r^{-1}\left(1-\mu(x)\right)$
and for  $\tilde \rho_r(x)=rx^{r-1}$  and $r>0$ we have 
$\varphi_r(x)=(1-x)^{1/r}$,
which transforms the full tent map into the
family (\ref{eq:fam1}).
Since 
families (\ref{eq:fam1}) and (\ref{eq:fam1a}) are topologically conjugated to the full tent map,
they are also conjugated to each other. In fact, the diffeomorphism conjugating them is
$\varphi_r(x)=\varphi_r^{-1}(x)=\left(1-x^r\right)^{1/r}$. This a nice example of two topologically conjugated maps 
sharing not only the Lyapunov exponent but also the natural
invariant measure.

This simple example shows that the maps found in previous
sections are not necessarily the only solution to the corresponding problem.
It was pointed out in Refs.\ \cite{Sogo1,Sogo2} that in general the inverse problem
does not have a unique solution and several examples were described with different maps
sharing a given invariant density while having different Lyapunov exponents in the form $\lambda_m=\ln m$.
We see here that it is also possible to have different maps with the 
same Lyapunov exponent and invariant measure.

An obvious variant of the method in this section is a starting family of densities $\rho_r(x)$ depending
on the parameter $r$ to construct a family of chaotic maps with a constant density
$\tilde\rho(x)$ by solving
\begin{equation}\label{eq:eqdiftil2}
 \tilde\rho\left(\varphi_r(x)\right)\left|\varphi'_r(x)\right|=\rho_r(x).
\end{equation}

\section{\label{sec:inverse}The inverse problem}

Obviously, one can combine the methods in the previous sections to get
families of maps depending on several parameters which display robust chaos.
Instead, we will discuss a more general and easier method to accomplish the same.

All the maps considered before can be written in the form $f=\mu^{-1}\circ g\circ\mu$, for
some appropriate $g$ and $\mu$. In fact, for the particular case of smooth 
unimodal families with robust chaos it has been proven that all 
maps within the family are topologically conjugate \cite{strien}. 
 This suggests a general way to construct a map $f$ (smooth or not, unimodal or not) with
prescribed Lyapunov exponent $\lambda$ and natural invariant density $\rho(x)$,
such that $\mu(x)\equiv\int_0^x\rho(y)\,dy$ is a diffeomorphism, even in the
case in which $f$, $\lambda$ and $\rho$ depend on one or several parameters, which
will not be written explicitly.

The starting point is a set of $m$ values $0<a_k<1$ such that
\begin{equation}\label{eq:suma}
b_m\equiv\sum_{k=1}^ma_i=1.
\end{equation}
If $b_0\equiv0$ and $b_k\equiv b_{k-1}+a_k$, for $k=1,2\ldots,m$,
the piecewise linear map
\begin{equation}\label{eq:mapg}
g(x)\equiv
\begin{cases}
\dfrac{x-b_{2i-2}}{a_{2i-1}},&b_{2i-2}\le x\le b_{2i-1};\\
\dfrac{b_{2i}-x}{a_{2i}},&b_{2i-1} \le x\le b_{2i}
\end{cases}
\end{equation}
(with $i=1,2,\ldots$) has the invariant density $\rho_0(x)=1$, because of (\ref{eq:suma}), and the
Lyapunov exponent
\begin{equation}\label{eq:lyapunovg}
0<\lambda_0=-\sum_{k=1}^m{a_k\ln a_k}\le \ln m.
\end{equation}
The maximum Lyapunov exponent $\lambda_0=\ln m$ is obtained with $a_1=a_2=\cdots=a_m=1/m$.

Now, if one chooses a value of $m$ large enough (so that
$\lambda\le\ln m$) and a set of
values $a_k$ (depending on the parameters of the desired map) such
that $\lambda=\lambda_0$, then the map $f=\mu^{-1}\circ g\circ\mu$
will have the prescribed Lyapunov exponent $\lambda$ and 
natural invariant density $\rho(x)=\mu'(x)$. The derivative $f'$ will be continuous
 at the maxima (minima) if $\lim_{x\to1}=\infty$ ($\lim_{x\to0}=\infty$).
The map $f$ will not be unique, since many others can be constructed in the same way with
higher values of $m$ and, even for the same $m$, there will be in general many (infinite)
ways of choosing the values $a_k$.
For instance, this method reduces to (\ref{eq:solmu1b}) for $m=2$, $a_1=a_2=1/2$
and $g(x)=T(x)=1-\left|1-2x\right|$. On the other hand, solution (\ref{eq:solmu2}) is recovered
with $m=2$, $a_2=1-a_1=1/(1+r)$ and $g$ given by (\ref{eq:fam1b}).

It is also easy to change slightly this method to construct maps with known 
Lyapunov exponent and piecewise natural density. Let us consider a single case.
Instead of (\ref{eq:mapg}) the starting point will be 
\begin{equation}\label{eq:mapgt}
\hat g(x)=g(x)
\begin{cases}
1,&0\le x\le b_{m-1};\\
a_1,&b_{m-1}\le x\le 1,
\end{cases}
\end{equation}
for $m=3,5,\ldots$ Since the orbit of $\tilde g(1)$ is $\mathcal{O}=\left\{a_1,1,a_1,\ldots\right\}$
there will be a single discontinuity point in the natural invariant density, which in fact 
is
\begin{equation}
\hat \rho_0(x)=\frac1{a_1(1+a_m)}
\begin{cases}
a_1+a_m,&0<x<a_1;\\
a_1,&a_1<x<1,
\end{cases}
\end{equation}
while the Lyapunov exponent is
\begin{equation}
0<\hat \lambda_0=-\frac{\displaystyle\sum_{k=1}^m{a_k\ln a_k}}{1+a_m} \le \mathrm{arcsinh}\, n,\quad  n\equiv(m-1)/2.
\end{equation}
The maximum value is reached with $a_1=\cdots=a_{m-1}=\sqrt{a_m}= \sqrt{1+n^2}-n$.
Then for each choice of the values $a_k$, the map
$\hat f=\mu^{-1}\circ\hat g\circ\mu$ will have the Lyapunov exponent $\hat\lambda=\hat\lambda_0$
and the natural invariant density
\begin{equation}            
 \hat \rho(x)=\frac{\rho(x)}{a_1(1+a_m)}
\begin{cases}
a_1+a_m,&0<x<\mu^{-1}\left(a_1\right);\\
a_1,&\mu^{-1}\left(a_1\right)<x<1.
\end{cases}
\end{equation}
For instance, taking $m=3$ and $a_1=a_2=\sqrt2-1$, one recovers the map (\ref{eq:saw}) for $r=\sqrt2+1$,
which can be extended to a piecewise linear family with known Lyapunov exponent and invariant
measure by letting $a_2$ run from $0$ to $2-\sqrt2$. In turn, applying to that family 
the diffeomorphism
$\mu(x)=\frac2\pi\arcsin\sqrt x$ one gets a family of smooth maps with known Lyapunov
exponent and piecewise continuous natural invariant density, which
coincides with (\ref{eq:fam3}) when $a_2=1/r=\sqrt2-1$.

It is easy to construct other examples starting from a piecewise linear
map with simple orbits of $\tilde g(1)$ (or $\tilde g(0)$).

\section{\label{sec:final}Final comments}

We have presented some new easy methods to construct families of one-dimensional
maps with robust chaos and exactly known Lyapunov exponent and natural invariant measure. 
As far as we know, this
is the first time general methods to do that are discussed. 

It should be stressed that 
the explicit examples with robust chaos discussed above have been selected for simplicity, but many
other can be easily constructed with the ideas discussed in this work.

\section*{Acknowledgments}
I am indebted to Prof.\ Kiyoshi Sogo for sending me a reprint of Ref. \cite{Sogo1}.
This work was supported by The University of the Basque Country
(Research Grant~GIU06/37).


\ifcase \dopreprint
\end{document}
\ifelse
\fi

\clearpage

\begin{figure}
\begin{center}
\includegraphics[width=\fwidth]{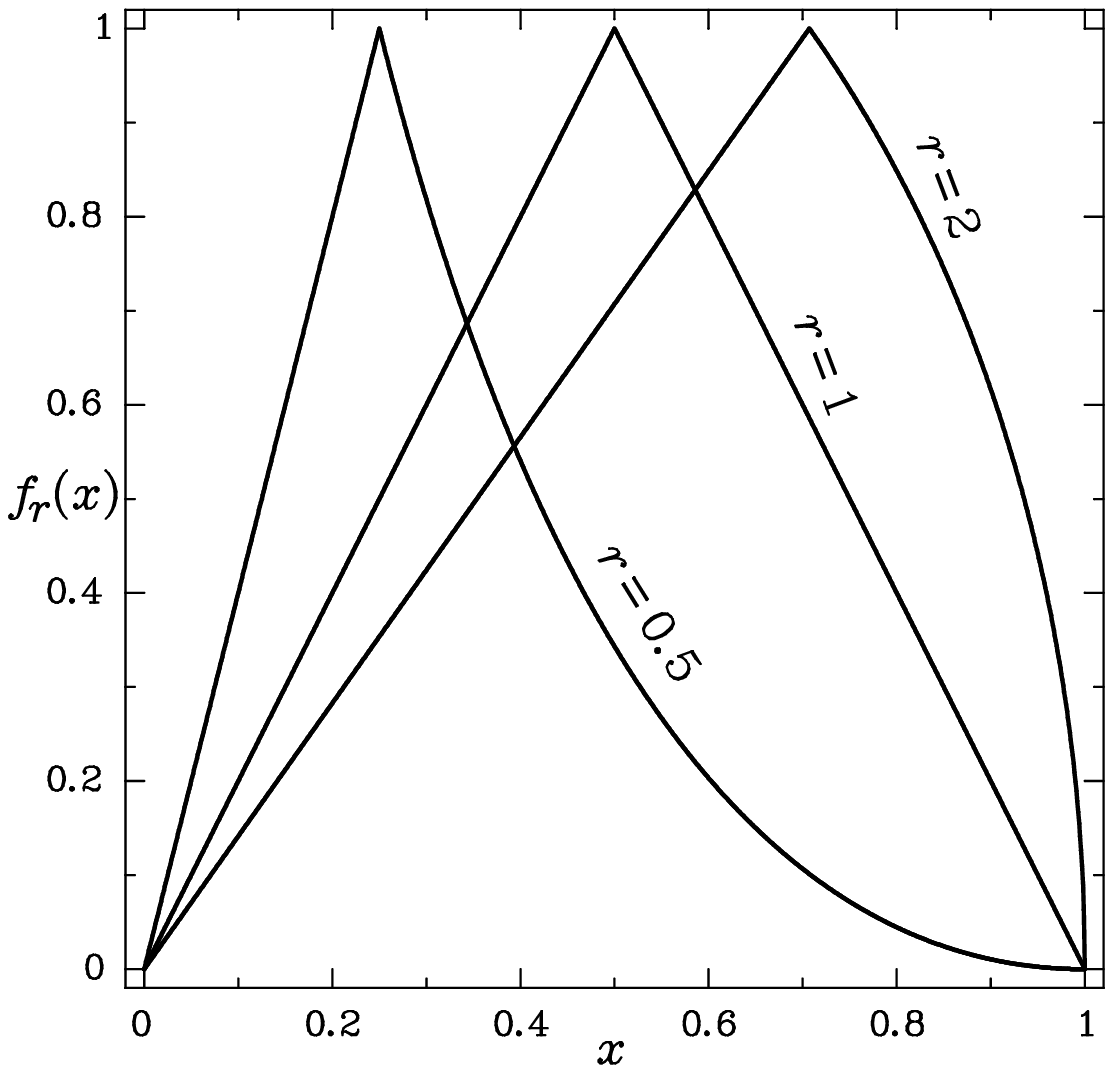}
\end{center}
\caption{Four members of the family  of maps (\ref{eq:fam1}).\label{fig1}} 
\end{figure}

\begin{figure}
\begin{center}
\includegraphics[width=\fwidth]{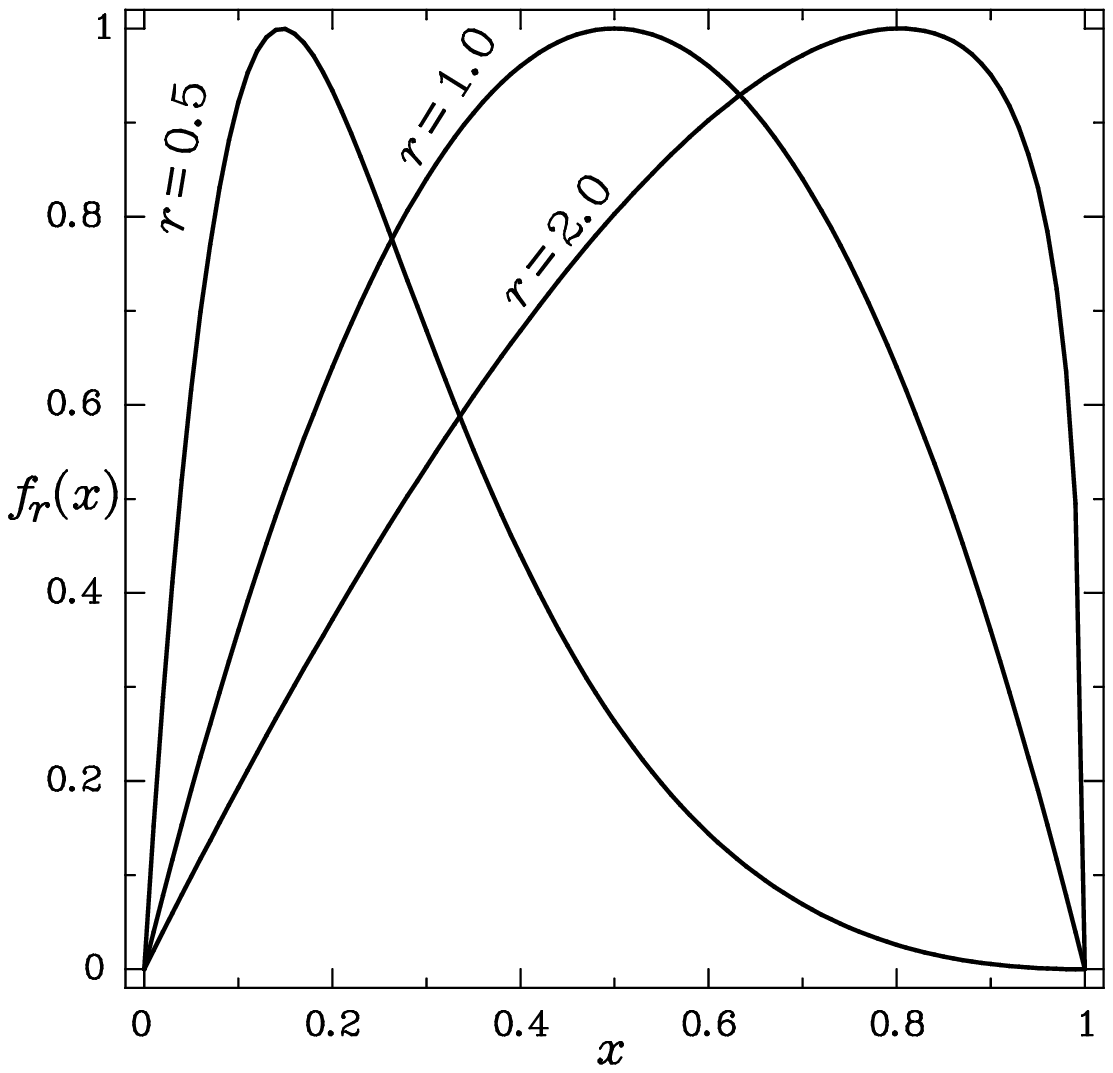}
\end{center}
\caption{Four members of the family of maps (\ref{eq:fam2}).\label{fig2}} 
\end{figure}

\begin{figure}
\begin{center}
\includegraphics[width=\fwidth]{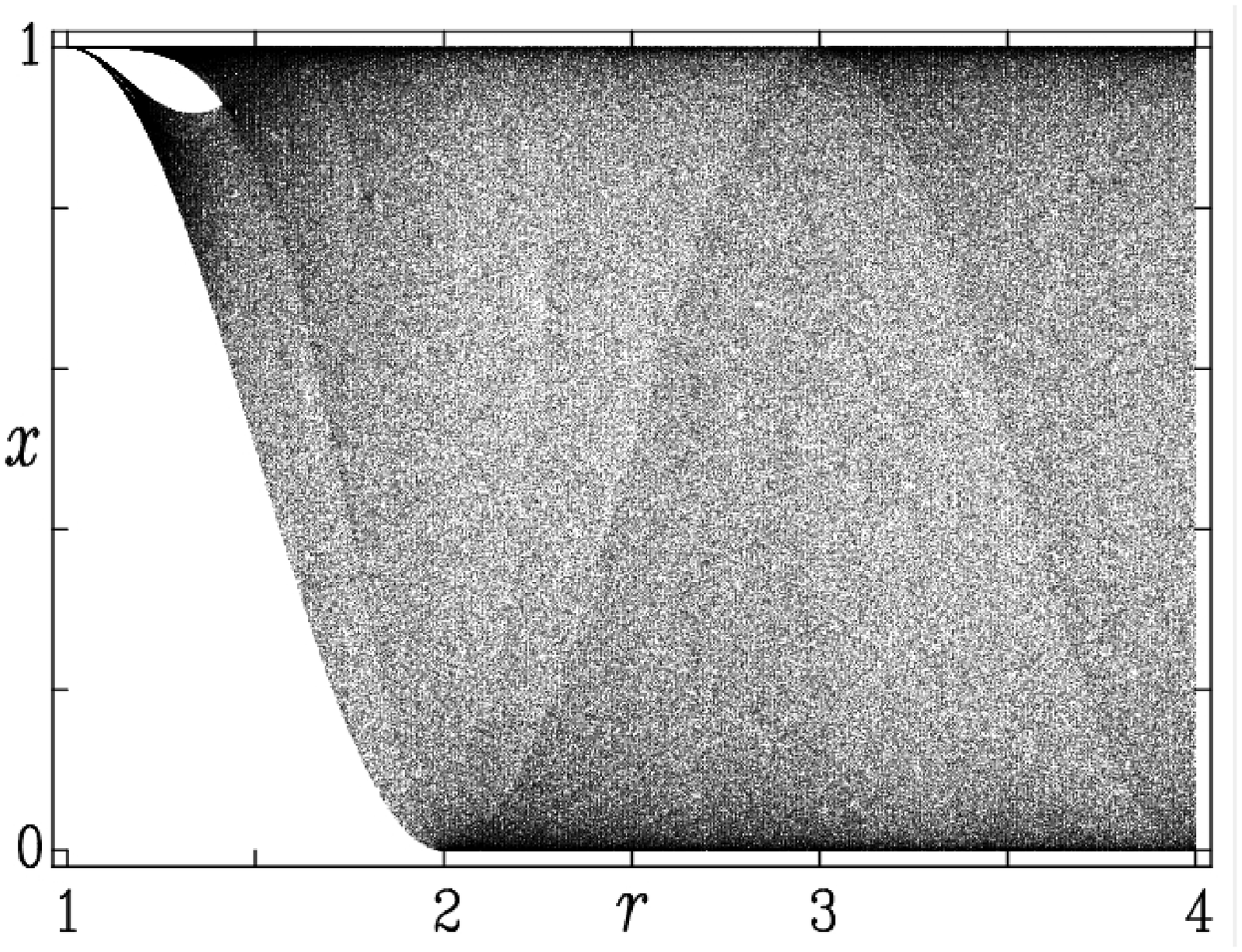}
\end{center}
\caption{Bifurcation diagram of the map (\ref{eq:fam3}).\label{fig3}} 
\end{figure}

\begin{figure}
\begin{center}
\includegraphics[width=\textwidth]{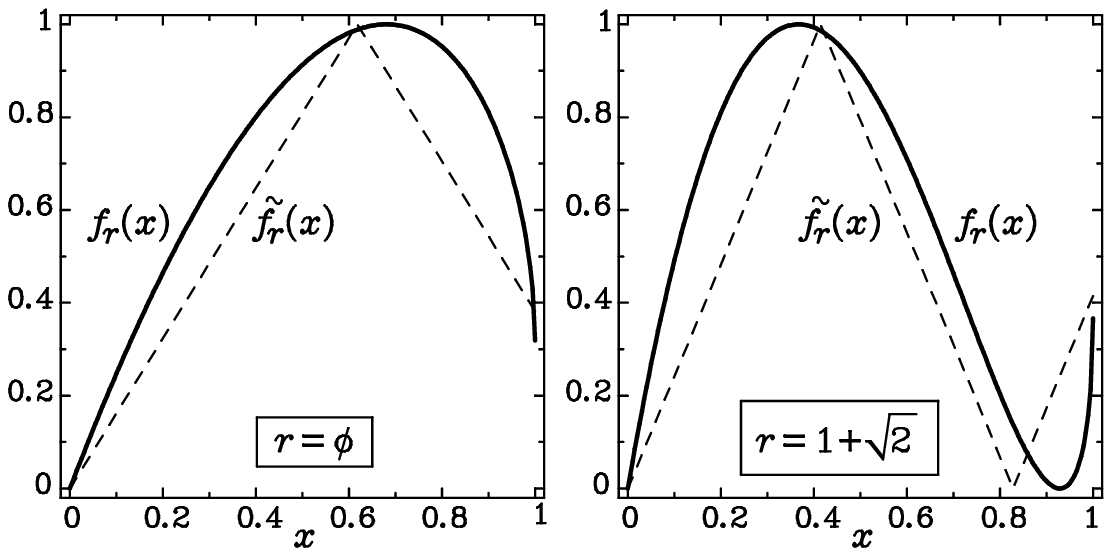}
\end{center}
\caption{The maps (\ref{eq:fam3})  and (\ref{eq:saw}) for $r=\phi,\ 1+\sqrt2$.\label{fig4}} 
\end{figure}

\end{document}

